# Quantum mechanical description of waveguides*

WANG Zhi-Yong[1†], XIONG Cai-Dong[1], HE Bing[2]

[1]*School of Optoelectronic Information, University of Electronic Science and Technology of China, Chengdu 610054, CHINA*
[2]*Department of Physics and Astronomy, Hunter College of the City University of New York, 695 Park Avenue, New York, NY 10021, USA*

In this paper, applying the spinor representation of the electromagnetic field, we present a quantum-mechanical description of waveguides. As an example of application, a potential qubit generated by photon tunneling is discussed.

## 1. Introduction

Waveguide theory is usually based on the classical or quantum field theory of the electromagnetic field, while in the first-quantized sense, the quantum-mechanical treatment of waveguides is absent. On the other hand, contrary to mechanical waves, the two-slit interference experiment of single photons shows that the behavior of classical electromagnetic waves corresponds to the quantum mechanical one of single photons, which is also different from the quantum-field-theory behavior such as the creations and annihilations of photons, the vacuum fluctuations, etc. In fact, in spite of the traditional conclusion that single photon cannot be localized,[1, 2] there have been developed photon wave mechanics which plays the role of the first-quantized theory of single photons,[3-10] and some recent studies have shown that photons can be localized in space.[11-13]

* Project supported by the National Natural Science Foundation of China (Grant No. 60671030) and by the Scientific Research Foundation for the Introduced Talents, UESTC (Grant No. Y02002010601022).
† Corresponding author.   E-mail: zywang@uestc.edu.cn

In this paper, in terms of the spinor representation of the electromagnetic field, we rewrite the Maxwell equations of free electromagnetic fields as the quantum-mechanical equation of single photons, and basing on which we will present a quantum-mechanical description of waveduides.

In the following, the natural units of measurement ($\hbar = c = 1$) is applied, repeated indices must be summed according to the Einstein rule, and the four-dimensional (4D) space-time metric tensor is chosen as $g^{\mu\nu} = \text{diag}(1,-1,-1,-1)$, $\mu,\nu = 0,1,2,3$. For our convenience, let $x^\mu = (t,-\boldsymbol{x})$, instead of $x^\mu = (t,\boldsymbol{x})$, denote the contravariant position four-vector (and so on), and then in our case $\hat{p}_\mu = i\partial/\partial x^\mu \equiv i\partial_\mu = i(\partial_t, -\nabla)$ denote 4D momentum operators.

## 2. Dirac-like equation of free photons

In vacuum the electric field, $\boldsymbol{E} = (E_1, E_2, E_3)$, and the magnetic field, $\boldsymbol{B} = (B_1, B_2, B_3)$, satisfy the Maxwell equations ($\hbar = c = 1$)

$$\nabla \times \boldsymbol{E} = -\partial_t \boldsymbol{B}, \quad \nabla \times \boldsymbol{B} = \partial_t \boldsymbol{E}, \tag{1}$$

$$\nabla \cdot \boldsymbol{E} = 0, \quad \nabla \cdot \boldsymbol{B} = 0. \tag{2}$$

Let $k_\mu = (\omega, \boldsymbol{k})$ denote the 4D momentum of photons ($\hbar = c = 1$), where $\omega$ is also the frequency and $\boldsymbol{k}$ the wave-number vector. The vectors $\boldsymbol{E}$ and $\boldsymbol{B}$ can also be expressed as the column-matrix form: $\boldsymbol{E} = (E_1\ E_2\ E_3)^\text{T}$ and $\boldsymbol{B} = (B_1\ B_2\ B_3)^\text{T}$ (the superscript T denotes the matrix transpose), by them one can define a 6×1 spinor $\psi(x)$ as follows:

$$\psi = \frac{1}{\sqrt{2}} \begin{pmatrix} \boldsymbol{E} \\ i\boldsymbol{B} \end{pmatrix}. \tag{3}$$

Moreover, by means of the 3×3 unit matrix $I_{3\times 3}$ and the matrix vector $\boldsymbol{\tau} = (\tau_1, \tau_2, \tau_3)$ with the components

$$\tau_1 = \begin{pmatrix} 0 & 0 & 0 \\ 0 & 0 & -i \\ 0 & i & 0 \end{pmatrix}, \quad \tau_2 = \begin{pmatrix} 0 & 0 & i \\ 0 & 0 & 0 \\ -i & 0 & 0 \end{pmatrix}, \quad \tau_3 = \begin{pmatrix} 0 & -i & 0 \\ i & 0 & 0 \\ 0 & 0 & 0 \end{pmatrix}, \tag{4}$$

we define the matrix $\beta_\mu = (\beta_0, \boldsymbol{\beta})$ or $\beta^\mu = (\beta^0, -\boldsymbol{\beta})$ ($\mu = 0,1,2,3$), where

$$\beta_0 = \beta^0 \equiv \begin{pmatrix} I_{3\times 3} & 0 \\ 0 & -I_{3\times 3} \end{pmatrix}, \quad \boldsymbol{\beta} \equiv (\beta_1, \beta_2, \beta_3) \equiv \begin{pmatrix} 0 & \boldsymbol{\tau} \\ -\boldsymbol{\tau} & 0 \end{pmatrix}, \tag{5}$$

Using Eqs. (3)-(5) one can rewrite the Maxwell equations as a Dirac-like equation

$$i\beta^\mu \partial_\mu \psi(x) = 0, \text{ or } i\partial_t \psi(x) = \hat{H}\psi(x), \tag{6}$$

where $\hat{H} = -i\beta_0 \boldsymbol{\beta} \cdot \nabla$ represents the Hamiltonian of free photons. In fact, one can show that the 6×1 spinor $\psi(x)$ transforms according to the $(1,0) \oplus (0,1)$ representation of the Lorentz group. Moreover, can prove that $(\beta^\mu \partial_\mu)(\beta_\nu \partial^\nu) = \partial^\mu \partial_\mu + \Omega$, where $\Omega \psi(x) = 0$ is identical with the transverse conditions given by Eq. (2), thus Eq. (6) implies that the wave equation $\partial^\mu \partial_\mu \psi(x) = 0$, where

$$\Omega = I_{2\times 2} \otimes \left[ \begin{pmatrix} \nabla_1 \\ \nabla_2 \\ \nabla_3 \end{pmatrix} \begin{pmatrix} \nabla_1 & \nabla_2 & \nabla_3 \end{pmatrix} \right], \tag{7}$$

The Dirac-like equation (6) is valid for all kinds of electromagnetic fields outside a source, and in the first-quantized sense it represents the quantum-mechanical equation of free photons. A more detailed discussion, see Ref. [14].

## 3. Analogy between guided waves and de Broglie matter waves

Though a TEM mode (its electric and magnetic fields are both perpendicular to the direction of propagation) cannot propagate in a single conductor transmission line, a guided wave can be viewed as the superposition of two sets of TEM waves being continually reflected back and forth between perfectly conducting walls and zigzagging down the waveguide, the two sets of TEM waves have the same amplitudes and frequencies, but reverse phases. Usually, the propagation of the electromagnetic wave through an ideal and uniform waveguide is described by a wave equation in (1+1) D space-time. However, in our formalism, we will generally place the waveguide along an arbitrary 3D spatial direction, by which we will show that guided waves have the same behaviors as de Broglie matter waves, and in terms of the 6×1 spinor defined by Eq. (5), we obtain a relativistic quantum equation for the guided waves.

In a Cartesian coordinate system spanned by an orthonormal basis $\{e_1, e_2, e_3\}$ with $e_3 = e_1 \times e_2$, we assume that a hollow metallic waveguide is placed along the direction of $e_3$, and the waveguide is a straight rectangular pipe with the transversal dimensions $b_1$ and $b_2$, let $b_1 > b_2$ without loss of generality. It is also assumed that the waveguide is infinitely long and its conductivity is infinite, and the electromagnetic source is localized at infinity. In the Cartesian coordinate system $\{e_1, e_2, e_3\}$, let $k_\mu = (\omega, \boldsymbol{k})$ denote the 4D momentum of photons inside the waveguide, then the wave-number vector is $\boldsymbol{k} = \sum_i e_i k_i = (k_1, k_2, k_3)$ and the frequency $\omega = |\boldsymbol{k}|$, where $k_1 = n\pi/b_1$ and $k_2 = s\pi/b_2$ ($n = 1, 2, 3...$, $s = 0, 1, 2...$), and the cutoff frequency of the waveguide is $\omega_{crs} = \sqrt{k_1^2 + k_2^2} = \pi\sqrt{(n/b_1)^2 + (s/b_2)^2}$. For simplicity, we shall restrict our discussion to the lowest-order cutoff frequency $\omega_c = \pi/b_1$. We define the effective rest mass of photons inside the waveguide as $m = \omega_c$, then the photon energy $E = \omega$ satisfies $E^2 = k_3^2 + m^2$. To obtain a Lorentz covariant formulation, let us rechoose a Cartesian coordinate system formed by an orthonormal basis $\{a_1, a_2, a_3\}$ with $a_3 = a_1 \times a_2$, such that in the new coordinate system, one has $e_3 k_3 = \sum_j a_j p_j = \boldsymbol{p}$. That is, in the new coordinate system, the waveguide is put along an arbitrary 3D spatial direction. Let $k_3 \geq 0$ without loss of generality. In the coordinate system $\{a_1, a_2, a_3\}$, one can read $\boldsymbol{k} = \boldsymbol{k}_\perp + \boldsymbol{k}_\parallel$, where (for the lowest-order mode $k_2 = 0$ and $k_1 = \omega_c$)

$$\boldsymbol{k}_\perp \equiv \boldsymbol{e}_1 k_1 + \boldsymbol{e}_2 k_2 = \boldsymbol{e}_1 \omega_c, \quad \boldsymbol{k}_\parallel = \boldsymbol{p} = \sum_i \boldsymbol{a}_i p_i = \boldsymbol{e}_3 k_3, \qquad (8)$$

stand for $\boldsymbol{k}$'s components being perpendicular and parallel to the waveguide, respectively. Furthermore, one can write the dispersion relation of photons inside the waveguide as $E^2 = p^2 + m^2$, it has the same form as the relativistic dispersion relation of free massive particles, where the cutoff frequency $\omega_c = m = |\boldsymbol{k}_\perp|$ plays the role of rest mass, while $\boldsymbol{p}$ represents the momentum of photons along the waveguide, such that the photons moving through the waveguide have an effective 4D momentum $p_{L\mu} \equiv (E, \boldsymbol{p})$.

According to the waveguide theory, the group velocity ($\boldsymbol{v}_g$) and phase velocity ($\boldsymbol{v}_p$) of photons along the waveguide are, respectively (note that $\hbar = c = 1$)

$$\begin{cases} \boldsymbol{v}_g = \boldsymbol{e}_3 \sqrt{1 - (\omega_c/\omega)^2} = \boldsymbol{p}/E \\ \boldsymbol{v}_p = \boldsymbol{e}_3 / \sqrt{1 - (\omega_c/\omega)^2} = \boldsymbol{e}_3 E/|\boldsymbol{p}| \end{cases}. \qquad (9)$$

Then one can obtain the following de Broglie's relations:

$$\begin{cases} \boldsymbol{v}_g \cdot \boldsymbol{v}_p = c^2 = 1 \\ \boldsymbol{p} = \hbar \boldsymbol{k}_\parallel = \boldsymbol{k}_\parallel \\ E = \hbar \omega = \omega = \sqrt{m^2 + p^2} \end{cases}. \qquad (10)$$

Using $m = \omega_c$ and Eqs. (9) and (10) one has

$$E = \frac{mc^2}{\sqrt{1 - (v_g^2/c^2)}} = \frac{m}{\sqrt{1 - v_g^2}}. \qquad (11)$$

This is exactly the relativistic energy formula. In fact, the group velocity $\boldsymbol{v}_g$ can be viewed as a relative velocity between an observer and a guided photon with the effective rest mass $m = \omega_c$. Eqs. (9)-(11) show that the behaviors of guided waves are the same as those of de Broglie matter waves, such that the guided photon can be treated as a free massive particle.

The effective rest mass of guided photons inside a hollow waveguide, as the *rest energy* of photons inside the waveguide (i.e., the energy as the group velocity $v_g = 0$), arises by forming standing-waves along the cross-section of the waveguide. In other words, it arises by freezing out the degree of freedom of transverse motion, or, by localizing and confining the electromagnetic energy along the cross-section of the waveguide. In terms of the effective rest mass $m = \hbar\omega_c/c^2 = \omega_c$, the effective Compton wavelength of guided photons is defined as

$$\lambdabar_c \equiv \hbar/mc = 1/\omega_c ,  \qquad (12)$$

As we know, it is impossible to localize a massive particle with a greater precision than its Compton wavelength, which is due to many-particle phenomena. Likewise, inside a hollow waveguide it is impossible to localize a photon along the cross section of the waveguide with a greater precision than its effective Compton wavelength, which owing to evanescent-wave phenomena (note that there always exists an inertial reference frame in which the frequency of a propagation mode is equal to the cutoff frequency of the waveguide).

## 4. Relativistic quantum-mechanical equation of guided photons

As we know, a light-like four-vector can be orthogonally decomposed as the sum of a space-like four-vector and a time-like four-vector. In our case, the time-like part of the light-like 4D momentum $k_\mu = (\omega, \boldsymbol{k})$ is the effective 4D momentum $p_{L,\mu} = (E, \boldsymbol{p})$, it represents the 4D momentum of photons moving along the waveguide, and is called the traveling-wave or active part of $k_\mu$; the space-like part of $k_\mu$ is the 4D momentum $p_{T\mu} \equiv (0, \boldsymbol{k}_\perp) = m\eta_\mu$ ($\eta_\mu \equiv (0, \boldsymbol{k}_\perp/m)$ satisfies $\eta_\mu \eta^\mu = -1$), it contributes to the effective rest mass, and is called the stationary-wave or frozen part of $k_\mu$. In other words, as the *rest energy* of photons inside the waveguide (i.e., the energy as the group velocity $v_g = 0$), the effective rest mass arises by freezing out the degrees of freedom related to the transverse motion of photons inside the waveguide. Therefore, we obtain an orthogonal decomposition for $k_\mu = (\omega, \boldsymbol{k})$ as follows:

$$k_\mu = (\omega, \boldsymbol{k}) = p_{T\mu} + p_{L\mu}, \quad p_{T\mu} \equiv (0, \boldsymbol{k}_\perp) = m\eta_\mu, \quad p_{L\mu} = (E, \boldsymbol{p}). \tag{13}$$

Such an orthogonal decomposition is Lorentz invariant because of $p_{L\mu} p_T^\mu = 0$.

Likewise, as for $x_\mu = (t, \boldsymbol{x})$ with $\boldsymbol{x} = \sum_i \boldsymbol{e}_i x_i = (x_1, x_2, x_3)$, in the coordinate system $\{\boldsymbol{a}_1, \boldsymbol{a}_2, \boldsymbol{a}_3\}$ one has $\boldsymbol{e}_3 x_3 = \sum_j \boldsymbol{a}_j r_j = \boldsymbol{r}$, i.e., the 3D vector $\boldsymbol{r}$ is parallel to the waveguide. Let $\boldsymbol{x}_\perp \equiv \boldsymbol{e}_1 x_1 + \boldsymbol{e}_2 x_2$, the orthogonal decomposition for $x_\mu$ can be written as

$$x_\mu = (t, \boldsymbol{x}) = x_{T\mu} + x_{L\mu}, \quad x_{T\mu} \equiv (0, \boldsymbol{x}_\perp), \quad x_{L\mu} = (t, \boldsymbol{r}). \tag{14}$$

It is easy to show that

$$k_\mu x^\mu = (p_{T\mu} + p_{L\mu})(x_T^\mu + x_L^\mu) = p_{T\mu} x_T^\mu + p_{L\mu} x_L^\mu. \tag{15}$$

The operator $\hat{p}_\mu = i\partial_\mu = i\partial/\partial x^\mu$ represents the totally 4D momentum operator of photons inside the waveguide, while

$$\hat{p}_{L\mu} = i\partial_{L\mu} = i\partial/\partial x_L^\mu, \tag{16}$$

represents the 4D momentum operator of photons moving along the waveguide. In spite of the boundary conditions for the waveguide, there are no charges in the free space inside the waveguide, and then where photons should obey the Dirac-like equation $i\beta^\mu \partial_\mu \psi(x) = 0$. Because of $\psi(x) \sim \exp(-ik_\mu x^\mu) = \exp[-i(p_{T\mu} x_T^\mu + p_{L\mu} x_L^\mu)]$, one has $p_{L\mu} \psi(x) = i\partial_{L\mu} \psi(x)$, and then

$$i\beta^\mu \partial_\mu \psi(x) = \beta^\mu k_\mu \psi(x) = \beta^\mu(p_{L\mu} + p_{T\mu})\psi(x) = \beta^\mu(i\partial_{L\mu} + p_{T\mu})\psi(x). \tag{17}$$

For a given waveguide and a given mode, $k_1 = n\pi/b_1$ and $k_2 = s\pi/b_2$ ($n = 1, 2, 3...$, $s = 0, 1, 2...$) are fixed, and then $p_{T\mu} = m\eta_\mu = (0, k_1, k_2, 0)$ is fixed. Using $i\beta^\mu \partial_\mu \psi(x) = 0$ and $p_{T\mu} = m\eta_\mu$, from Eq. (17) one can obtain $i\beta^\mu(\partial_{L\mu} - im\eta_\mu)\psi(x) = 0$. Let $\psi(x) = \varphi(x_L)\exp(-ip_{T\mu}x_T^\mu)$, obviously $\varphi(x_L) \sim \exp(-ip_{L\mu}x_L^\mu)$, we obtain the Dirac-like equation of photons moving along the waveguide

$$i\beta^\mu(\partial_{L\mu} - im\eta_\mu)\varphi(x_L) = 0. \tag{18}$$

Using Eq. (18) and $\eta^\mu \partial_{L\mu}\varphi(x_L) = \partial_{L\mu}\eta^\mu \varphi(x_L) = 0$ one can obtain the Klein-Gordon equation

$$(\partial_{L\mu}\partial_L^\mu + m^2)\varphi(x_L) = 0. \tag{19}$$

In the first-quantized sense, Eqs. (18) and (19) serve as the relativistic quantum-mechanical equations of photons moving along the waveguide, while in the second-quantized sense, they are quantum-field-theory equations.

Eqs. (18) and (19) are expressed in the arbitrary coordinate system $\{a_1, a_2, a_3\}$ (as viewed in which the waveguide is put along an arbitrary 3D spatial direction), they can be simplified in the coordinate system $\{e_1, e_2, e_3\}$ (as viewed in which the waveguide is put along the $x_3$-axis, and then one has $x_{L\mu} = (t, 0, 0, x_3)$ and $m\eta_\mu = (0, k_1, k_2, 0)$). To be specific, in the coordinate system $\{e_1, e_2, e_3\}$, Eq. (18) becomes

$$(i\beta^0 \partial_t + i\beta^3 \partial_3 + \beta^1 k_1 + \beta^2 k_2)\varphi(t, x_3) = 0. \tag{20}$$

Let $\psi(t, \boldsymbol{x}) = \varphi(t, x_3)\exp[-i(k_1 x^1 + k_2 x^2)]$, one can easily derive Eq. (6) from Eq. (20), just as one expected. Likewise, let $\varphi(x_L) = \varphi(t, x_3) = \exp(i\omega t)\phi(x_3)$ and $r = x_3$, Eq. (19) becomes the usual form

$$[\partial^2/\partial r^2 + (\omega^2 - \omega_c^2)]\phi(r) = 0. \tag{21}$$

As mentioned before, for the lowest order mode one has $k_1 = \omega_c$ and $k_2 = 0$, and then using $r = x_3$ Eq. (20) can be rewritten as the Schrödinger equation (note that $\beta^0 = \beta_0$ while $\beta^j = -\beta_j$ ($j=1,2,3$))

$$i\frac{\partial}{\partial t}\varphi(t,r) = (i\beta_0\beta_3\frac{\partial}{\partial r} + \beta_0\beta_1\omega_c)\varphi(t,r). \tag{22}$$

For the time being, the term of $\beta_0\beta_1\omega_c$ in Eq. (22) can also be regarded as a potential.

## 5. A potential qubit via photon tunneling

As we know, the simplest elementary building blocks for a quantum computer are quantum bits (qubits), i.e., two-level quantum systems. [15-18] In the following we will discuss a two-level quantum system induced by photon tunneling.

For convenience, from now on we will discuss Eqs. (18) and (19) in the coordinate system $\{e_1, e_2, e_3\}$, in which Eq. (18) becomes Eq. (22) while Eq. (19) becomes Eq. (21). In the first-quantized sense, by replacing $\varphi(t,r)$ with the state vector $|t\rangle$, Eq. (22) becomes the following quantum-mechanical equation:

$$i\frac{\partial}{\partial t}|t\rangle = [\hat{H}_0 + V(r)]|t\rangle, \tag{23}$$

where $\hat{H}_0 = i\beta_0\beta_3\partial/\partial r$ is regarded as the free part of the Hamiltonian $\hat{H} = \hat{H}_0 + V(r)$, while $V(r) = \beta_0\beta_3\omega_c(r)$ plays the role of a potential matrix (its eigenvalues are potentials). Here, we have rewritten the cut-off frequency $\omega_c$ as the function $\omega_c(r)$ of the coordinate $r = x_3$ along the waveguide. Applying Eqs. (4) and (5), one can prove that the eigenvalues of $\beta_0\beta_3$ are $\lambda_1 = \lambda_2 = 1$, $\lambda_3 = \lambda_4 = -1$, and $\lambda_5 = \lambda_6 = 0$, where $\lambda_{1,2}$ and $\lambda_{3,4}$ are related to the solutions of the transverse photons while $\lambda_{5,6}$ to the solutions of longitudinal and scalar photons. Here do not involve any non-radiation field, then we just consider the solutions of $\lambda_{1,2}=1$ and $\lambda_{3,4}=-1$. As for our issue, the two solutions will present us with the same conclusion, then for convenience we will only concern the eigenvalues of $\lambda_1=\lambda_2=1$, for the moment the matrix $\beta_0\beta_3$ can be replaced with $\lambda_1=\lambda_2=1$ directly. For our purpose, assume that the waveguide extends from $r=-a$ to $r=a$ with two closed ends at $r=\pm a$, and has a variable cross-section such that the cut-off frequency $\omega_c=\omega_c(r)$ satisfies:

$$\omega_c(r)=\begin{cases}+\infty,\text{ for }|r|\geq a\\ \theta\ll\varepsilon,\text{ for }a\geq|r|\geq b,\\ \varepsilon,\text{ for }b\geq|r|\geq 0\end{cases} \quad (24)$$

where $\varepsilon>0$ is a real constant. Physical observables involve energy differences and not the absolute value of the energy, and do not depend on the choice of zero-energy reference point. For simplicity, as $a\geq|r|\geq b$, we choose $V(r)=\beta_0\beta_3\theta$ ($\ll\beta_0\beta_3\varepsilon$) as the zero-energy reference point, and then the potential can be written as

$$V(r)=\begin{cases}+\infty,\text{ for }|r|\geq a\\ 0,\text{ for }a\geq|r|\geq b\\ V=\beta_0\beta_3\varepsilon,\text{ for }b\geq|r|\geq 0\end{cases}. \quad (25)$$

In other words, within the region $a\geq|r|\geq b$, the size of section of the waveguide is approximatively regarded as an infinite one. Obviously, $V(r)=\beta_0\beta_3\varepsilon$ is commutative with $\hat{H}_0=i\beta_0\beta_3\partial/\partial r$, such that they have common eigenstates. For the eigenvalues $\lambda_1=\lambda_2=1$ of $\beta_0\beta_3$, the potential $V(r)$ possesses two minima ($V(r)=0$) at the intervals $a\geq r\geq b$ and $-b\geq r\geq -a$, respectively. That is, Eq. (25) corresponds to a double-minimum potential problem which is usually discussed by taking the ammonia molecule as an example.[19] For the moment, the eigenvalues of $V=\beta_0\beta_3\varepsilon$ are $\varepsilon>0$, and for our purpose, we assume that the eigenvalues of $\hat{H}_0$ are $E_0<\varepsilon$. That is, within the region $a\geq|r|\geq b$, the frequency of a photon inside the waveguide is smaller than the cut-off frequency of the waveguide for $b \geq r \geq -b$, and then with respect to the photon lying in $a \geq |r| \geq b$, the waveguide for $b \geq r \geq -b$ is an undersized waveguide.

Applying Eq. (25), within the region $a \geq |r| \geq b$, Eq. (23) becomes

$$i\frac{\partial}{\partial t}|t\rangle = \hat{H}_0 |t\rangle, \qquad (26)$$

Inside the waveguide, as a photon lies in the region $-b \geq r \geq -a$, let $|1\rangle$ denote its quantum state; as the photon lies in the region $a \geq r \geq b$, let $|2\rangle$ denote its another quantum state. The regions $-b \geq r \geq -a$ and $a \geq r \geq b$ are two alternative positions of the photon, where the photon has the same energy $E_0$. The quantum states $|1\rangle$ and $|2\rangle$ together forms a complete base set: $\sum_i |i\rangle\langle i| = I_{2\times 2}$ ( $i = 1,2$ ), by which one has $|t\rangle = \sum_i C_i |i\rangle$, and Eq. (26) can be written as

$$i\frac{\partial}{\partial t}\begin{pmatrix} C_1 \\ C_2 \end{pmatrix} = \begin{pmatrix} H_{11} & H_{12} \\ H_{21} & H_{22} \end{pmatrix}\begin{pmatrix} C_1 \\ C_2 \end{pmatrix}, \qquad (27)$$

where

$$H_{ij} = \langle i|\hat{H}_0|j\rangle, \quad C_i = \langle i|t\rangle, \quad i,j = 1,2. \qquad (28)$$

From the point of view of classical mechanics, the photon can not propagate through the undersized waveguide within $b \geq r \geq -b$, and the regions $-b \geq r \geq -a$ and $a \geq r \geq b$ are two separate ones, such that the quantum states $|1\rangle$ and $|2\rangle$ are orthogonal, and Eq. (26) becomes

$$i\frac{\partial}{\partial t}\begin{pmatrix} C_1 \\ C_2 \end{pmatrix} = \begin{pmatrix} E_0 & 0 \\ 0 & E_0 \end{pmatrix}\begin{pmatrix} C_1 \\ C_2 \end{pmatrix}. \qquad (29)$$

For the moment, the quantum states $|1\rangle$ and $|2\rangle$ are the eigenstates of the system with doubly degenerating.

However, from the point of view of quantum mechanics, because of photon tunneling through the undersized waveguide (for $b \geq r \geq -b$), the quantum states $|1\rangle$ and $|2\rangle$ are not orthogonal, that is, there is a coherent superposition between them, which implies that $H_{ij} \neq 0$ for $i \neq j$. Because $H_{ij}$'s are Hermitian, one can prove that $H_{ij} = H_{ji}$. Let $H_{12} = H_{21} = -A$, where $A$ is a positive real quantity, Eq. (27) becomes

$$\mathrm{i}\frac{\partial}{\partial t}\begin{pmatrix} C_1 \\ C_2 \end{pmatrix} = \begin{pmatrix} E_0 & -A \\ -A & E_0 \end{pmatrix}\begin{pmatrix} C_1 \\ C_2 \end{pmatrix}. \tag{30}$$

Let $|\pm\rangle = (|1\rangle \pm |2\rangle)/\sqrt{2}$, $C_\pm = \langle \pm | t \rangle$, one can prove that $C_\pm = (C_1 \pm C_2)/\sqrt{2}$, and

$$\mathrm{i}\frac{\partial}{\partial t}\begin{pmatrix} C_+ \\ C_- \end{pmatrix} = \begin{pmatrix} E_0 - A & 0 \\ 0 & E_0 + A \end{pmatrix}\begin{pmatrix} C_+ \\ C_- \end{pmatrix}. \tag{31}$$

Therefore, via photon tunneling, the quantum states $|+\rangle$ and $|-\rangle$ become the eigenstates of the system with an energy-level splitting of $2A$ between the eigenstates $|+\rangle$ and $|-\rangle$, and the original bistable system with double minimum potential now becomes a two-level quantum system with an unique ground state. Such a two-level quantum system as a qubit, may have potential application in future quantum computation and quantum information.

## 6. Conclusions

There exists a closely analogy between the behaviors of de Broglie matter waves and those of electromagnetic waves inside a waveguide, such that the guided photons can be treated as free massive particles, by which we present a quantum mechanical description of waveguides. As an example of application, a possible qubit induced by photon tunneling is presented.

## Acknowledgement

The first author (Wang Z Y) would like to thank professor Ole Keller for his helpful discussions.

References

[1] Newton T D and Wigner E P 1949 *Rev. Mod. Phys*. **21** 400.

[2] Wightman A S 1962 *Rev. Mod. Phys*. **34** 854-872.

[3] Mandel L 1966 *Phys. Rev*. **144** 1071.

[4] Cook R J 1982 *Phys. Rev. A* **25** 2164.

[5] Inagaki T 1994 *Phys. Rev. A* **49** 2839.

[6] Bialynicki-Birula I 1994 *Acta Phys. Polon. A* **86** 97.

[7] Bialynicki-Birula I 1996 *Progress in Optics XXXVI* edited by Wolf E (Amsterdam: Elsevier).

[8] Bialynicki-Birula I 1996 *Coherence and Quantum Optics* **VII** edited by Eberly J, Mandel L and Wolf E (New York: Plenum Press) p313.

[9] Sipe J E 1994 *Phys. Rev. A* **52** 1875.

[10] Keller O 2005 *Phys. Rep*. **411** 1.

[11] Adlard C, Pike E R and Sarkar S 1997 *Phys. Rev. Lett* **79** 1585.

[12] Bialynicki-Birula I 1998 *Phys. Rev. Lett* **80** 5247.

[13] Chan K W, Law C K, and Eberly J H 2002 *Phys. Rev. Lett* **88** 100402.

[14] Wang Z Y, Xiong C D and Keller O 2007 *Chin. Phys. Lett*. **24** 418.

[15] Beth T and Leuchs G 2005 *Quantum Information Processing* second edition (Weinheim: Wiley Verlag).

[16] Nielsen M A and Chuang I L 2000 *Quantum Computation and Quantum Information*, (Cambridge: Cambridge University Press).

[17] Wang Z Y and Xiong C D 2006 *Chinese Physics* **15** 2223.

[18] Jin L J and Fang M F 2006 *Chinese Physics* **15** 2012.

[19] Hecht K T 2000 *Quantum Mechanics* (New York: Springer-Verlag) pp45-48 and pp76-78.